\documentclass{jfm}

\def\p{\partial}
\def\be{\begin{eqnarray}}
\def\ee{\end{eqnarray}}
\def\bes{\begin{subeqnarray}}
\def\ees{\end{subeqnarray}}

\def\f{\frac}
\def\lp{\left(}
\def\rp{\right)}

\def\befi{\begin{figure}}
\def\eefi{\end{figure}}
\def\a{\alpha}

\def\bce{\begin{center}}
\def\ece{\end{center}}

\def\d{\delta}

\def\th{\theta}

\def\d{\textrm{d}}

\def\ba#1\ea{\begin{align}#1\end{align}}

\def\bsa#1\esa{\begin{subequations}
\begin{align}#1\end{align} \end{subequations}}

\usepackage{graphicx}
\usepackage{newtxtext}
\usepackage{newtxmath}
\usepackage{natbib}
\usepackage{hyperref}
\hypersetup{
    colorlinks = true,
    urlcolor   = blue,
    citecolor  = black,
}

\newcommand{\RomanNumeralCaps}[1]
\nolinenumbers

\shortauthor{T. Iida and A. Zareei and M.-R. Alam}
\shorttitle{Deep water wave cloaking using multi-layered plate}
\title{Deep water wave cloaking using a multi-layered buoyant plate}

\author{ Takahito Iida\aff{1}$^*$, Ahmad Zareei\aff{2}$^*$ \and Mohammad-Reza Alam\aff{3} \corresp{email: {reza.alam@berkeley.edu}; $^*$ T.I. and A.Z. contributed equally.}
 }

\affiliation{
\aff{1}
Dept. of Naval Architecture and Ocean Engineering, Osaka University, Osaka 5650871, Japan
\aff{2}
Harvard John A. Paulson School of Engineering and Applied Sciences, Harvard University, Cambridge, MA 02138, USA
\aff{3}
Dept. of Mechanical Engineering, U.C. Berkeley, Berkeley, CA 94720, USA
}

\begin{document}
\maketitle

\begin{abstract}
Trajectory of surface gravity waves in potential flow regime is affected by the gravitational acceleration, water density, and sea bed depth. While the gravitational acceleration and water density are approximately constant, and the effect of water depth on surface gravity waves exponentially decreases as the water depth increases. In shallow water, cloaking an object from surface waves by varying the sea bed topography is possible, however, as the water depth increases, cloaking becomes a challenge since there is no physical parameter to be engineered and subsequently affect the wave propagation. In order to create an omnidirectional cylindrical cloaking device for finite-depth/deep-water waves, we propose a multi-layered elastic plate that floats on the surface around a to-be-cloaked cylinder. The buoyant elastic plate is made of axisymmetric, homogeneous, and isotropic layers which provides adjustable degrees of freedom to engineer and affect the wave propagation. We first develop a pseudo-spectral method to efficiently determine the wave solution for a multi-layered buoyant plate. Next, we optimize the physical parameters of the buoyant plate (i.e. elasticity and mass of every layer) using a real-coded evolutionary algorithm to minimize the energy of scattered-waves from the object and therefore cloak the inner cylinder from incident waves. We show that the optimized cloak reduces the energy of scattered-waves as high as 99.2\% for the target wave number. We quantify the effectiveness of our cloak with different parameters of the buoyant plate and show that varying the flexural rigidity is essential to control wave propagation and the cloaking structure needs to be at least made of four layers with a radius of at least three times of the cloaked region. 
We quantify the wave drift force exerted on the structures and show that the buoyant plate reduces the exerted force by 99.9\%. The proposed cloak, due to its structural simplicity and effectiveness in reducing the wave drift force, may have potential applications in cloaking offshore structures from deep water waves.
\end{abstract}



\section{Introduction}
\label{intr}


{To safely operate and maintain offshore structures in deep water, such as column-based (spar-type) wind turbines or floating platforms, over a long period of time, the reduction of the wave drift force (time-averaged second-order hydrodynamic force) is essential. Additionally, creating a calm and cloaked region on the ocean surface in presence of surface waves facilitates the installation of these offshore structures. 
The concept of cloaking objects from incident waves was initially developed for electromagnetic waves~\cite[]{Pendry2006, Schurig2006}. A cloak is a structure that encloses a to-be-cloaked object and results in no reflection or scattering of incident waves from the object. The downstream waves bears no information of the object's presence and the cloaked region is secluded from incident waves. The cloaks proposed for the electromagnetic waves are based on the idea of transformation optics and exploit the form-invariance of governing equations. The material properties of the cloak are found using spatial transformations of governing equations. The obtained material properties are usually inhomogeneous and orthotropic and implementing such properties results in the desired trajectory of the incoming waves. Since the only property used in this technique is the form-invariance of the governing equations under spatial coordinate transformation, this method has been extended to other areas of physics with form-invariant governing equations, such as acoustics \cite[]{Cummer2007, Chen2007, Huang2014,zareei2018continuous,darabi2018broadband}, elastic waves \cite[]{Farhat2009, Stenger2012,darabi2018experimental,zareei2017broadband}, or seismic waves \cite[]{Brule2014}. Besides transformation techniques, an alternative method to achieve cloaking is to minimize the scattering cross-section (or energy of scattered-waves) of an object \cite[]{alu2005achieving}. This method has also been successfully applied to different types of waves such as acoustic waves \cite[]{guild2011cancellation} or water waves \cite[]{Porter2011, Porter2014}.}

{In shallow water, cloaking has been realized using the coordinate transformation technique~\cite[][]{Berraquero2013,Zareei2015}.
The governing equation of shallow water gravity waves is form-invariant under the coordinate transformation.
The physical parameters controlling wave propagation in shallow water are the sea bed topography and the gravitational acceleration \cite[e.g.][]{Berraquero2013,Zareei2015}. Clearly varying the gravitational acceleration is unrealistic. 
The use of \textit{nonlinear transformation} of cylindrical region can keep the gravitational acceleration a constant value, and only changes in the sea bed topography is used to achieve cloaking from shallow water waves \cite[]{Zareei2015}.
Alternative approaches are to use an array of bottom-mounted objects for implementing a cloaking devices for shallow water surface gravity waves using the coordinate transformation technique~\cite[]{Dupont2016, Iida2018}, or capillary-gravity waves~\cite[]{Farhat2008}.}

{As the water depth increases, the effect of the bottom topography exponentially decreases, and as a result engineered sea bed topography does not affect the wave propagation. 
In addition, the governing equation of deep water waves in potential flow regime is not form-invariant under the coordinate transformation, and therefore it is hard to apply the transformation technique to make a cloak for deep water waves.
To achieve cloaking in a finite depth, \textit{a scattering cancellation} is proposed \cite[]{Porter2011, Porter2014}. 
Specifically, a sea bed topography is designed to cancel the energy of scattered-waves of a bottom-mounted cylinder \cite[]{Porter2014}.
Multiple floating cylinders (or a ring) are used and this shows cloaking performance even for deep water waves \cite[]{Newman2014}.
This cloaking method is validated by numerical computations using a higher-order boundary element method \cite[]{Iida2014} and a model experiment \cite[]{Iida2016}. 
Since an asymmetrical array of the cylinder is utilized, the performance of the cloak depends on the wave direction \cite[]{Zhang2019}.
A concentric annular elastic plate is optimized to reduce the wave drift force \cite[]{Loukogeorgaki2019}, however, the reduction of the energy of scattered-waves is not observed since the plate's flexural rigidity is the only controllable medium property. 
Besides, since the flexural waves in elastic thin plates are not form-invariant \cite[]{zareei2017broadband}, the floating elastic plate cannot be used for the transformation-based cloaking in deep water.
} 

{In this paper, we implement a cloak using a multi-layered elastic plate floating on the surface surrounding a to-be-cloaked cylinder in order to develop a practical and realistic offshore structure protection device from waves. 
The multi-layered plate is made of concentric, axisymmetric, homogeneous, and isotropic elastic layers that provide adjustable degrees of freedom in physical parameters to vary and affect the wave propagation (See Fig. \ref{fig1}). Medium parameters of this plate, i.e. the size, the flexural rigidity, and the mass of each layer, are optimized to cancel the energy of scattered-waves from the cylinder. First, we develop a numerical scheme for solving this problem. 
The scheme is based on pseudo-spectral and eigenvalue matching methods \cite[]{Peter2004}, and this is extended to the calculation of the multi-layered plate.
Next, we use an evolutionary strategy to find the optimum parameters for the floating plate.
We quantify the effectiveness of our multi-layered cloak and show that the energy of scattered-waves and wave drift force acting on the structure are reduced as high as 99.2\% at the target wave number. In addition, we study the effect of different parameters on the optimum solution, and show that varying the flexural rigidity is essential to cloaking performance and the cloaking structure needs to be at least be made of four layers with a radius of at least three times of the cloaked cylinder. By quantifying the wave drift force exerted on the cylinder, we show that the buoyant plate reduces the net force by 99.9\% at the target wave number.}

\section{Formulation of the problem}
\subsection{Governing equation and boundary conditions}

\begin{figure}
 \centering
 \includegraphics[width=4.00in]{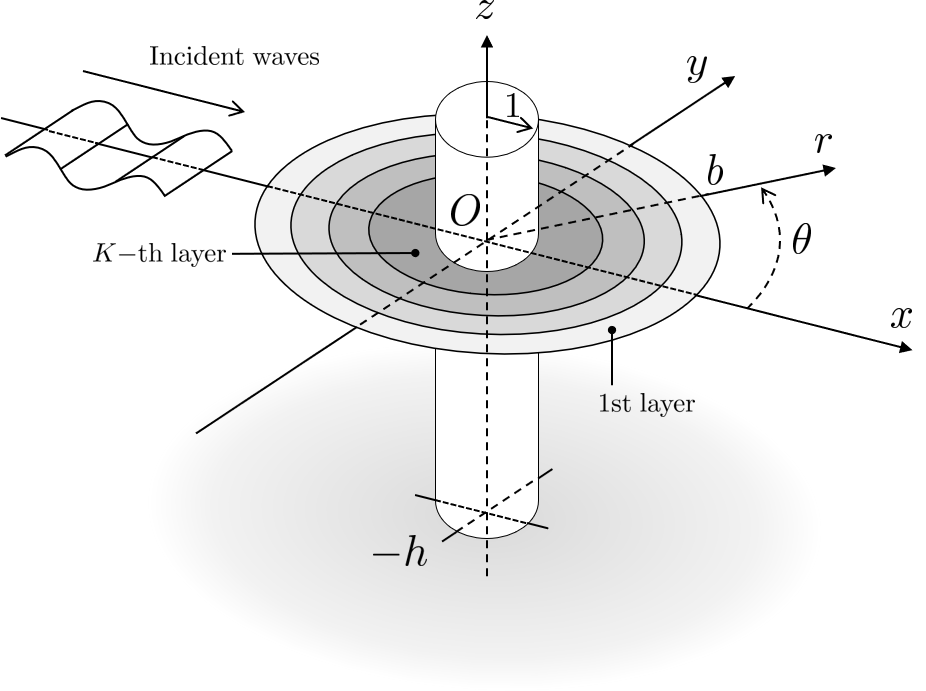}
 \caption{Schematic representation of bottom-mounted cylinder and multi-layered buoyant elastic plate. All values are normalized by radius of cylinder, wave amplitude, fluid density, and gravitational acceleration. Sea bottom is flat at $z=-h$. Plate consists of $K$ horizontally concentric annular layers, and outermost radius of plate is $b$. Waves are incident from negative $x$-direction.}
 \label{fig1}
\end{figure}

We consider a bottom-mounted cylinder with radius $a$ surrounded by a multi-layered circular elastic plate floating on the surface of the water which is expected to effectively cloak the cylinder from incident water waves (see. Fig. \ref{fig1}). The plate is made of $K$ horizontal layers where the mechanical properties of each layer is kept as a constant. This assumption facilitates future experimental validation of our setup. We number the layers from outer to inner and denote the outer/inner radius of the $k$-th layer as $R_\text{o}^{(k)}/R_\text{i}^{(k)}$. The outermost and innermost layers are respectively  $R_\text{o}^{(1)}=b$ and $R_\text{i}^{(K+1)}=a$. For simplicity of notation we will use $R^{(k)}$ for the outer radius unless mentioned other-wised. We consider a Cartesian coordinate system with origin $O$ at the center of the cylinder where $z=0$ plane coincides with the undisturbed free surface of the water and positive $z$-axis pointing upward. Assuming a flat bottom topography at $z=-h$ and considering that the surrounding fluid to be incompressible,  inviscid,  and irrotational, the linearized governing equation and boundary conditions for the fluid's velocity potential $\Phi(\mbox{\boldmath $x$}, t)$ become
\begin{eqnarray}
\vspace{1mm}
& \displaystyle \nabla^2\Phi=0 & -h\le z\le 0, \label{eqL}\\%
\vspace{1mm}
& \displaystyle \frac{\partial \Phi}{\partial z}=0 & z=-h, \label{eqB}\\%
\vspace{1mm}%
& \displaystyle g\frac{\partial \Phi}{\partial z}+ \frac{\partial^2 \Phi}{\partial t^2}=0 & z=0, r\ge b, \label{eqF}\\%
& \displaystyle \bigg(D^{(k)}\nabla_\perp^4+m_c^{(k)}\frac{\partial^2}{\partial t^2}+\rho_w g\bigg)\frac{\partial \Phi}{\partial z}+\frac{\partial ^2 \Phi}{\partial t^2}=0 \quad & z=0, R^{(k+1)}\le r < R^{(k)},\label{eqCk}%
\end{eqnarray}%
where $g$ is the gravitational acceleration, $\rho_w$ is the fluid density, $D^{(k)}$ is the flexural rigidity of the $k$-th layer, $m_c^{(k)}$ is mass per unit length of the $k$-th layer, $\nabla=(\partial/\partial_x, \partial/\partial_y, \partial/\partial_z)$ is the gradient operator, and  $\nabla_\perp = (\partial/\partial_x, \partial/\partial_y)$ is the horizontal gradient operator.  Flexural rigidity and mass of the cloaking plate are calculated as $D^{(k)}=E^{(k)}t_c^{(k)3}/12(1-\nu^2)$ and $m_c^{(k)}=\rho_c^{(k)} t_c^{(k)}$ where $E^{(k)}$ is the $k$-th layer Young's modulus, $\nu$ is the Poisson's ratio, and $\rho_c^{(k)}$ and $t_c^{(k)}$ are the density and thickness of the $k$-th layer, respectively. In the above equations, Eq. \eqref{eqL} is obtained by the mass conservation, Eq. \eqref{eqB} is a bottom kinematic boundary condition, Eq. \eqref{eqF} is a linearized free surface condition, and Eq. \eqref{eqCk} is the linearized surface condition for the $k$-th layer of the plate. Next, we non-dimensionalize the above equations, using the radius of the cylinder $a$, incident wave amplitude $\zeta_w$, fluid density $\rho_w$, and gravitational acceleration $g$, where we define $\bar{x}=x/a$, $\bar{t}=t\sqrt{g/a}$, and $\bar{\Phi}=\Phi/(\zeta_w\sqrt{ag})$.
We further assume time-harmonic solutions, the velocity potential becomes $\bar{\Phi}(\bar{\mbox{\boldmath $x$}}, \bar{t})={\rm Re}[\bar{\phi}(\bar{\mbox{\boldmath $x$}}){\rm exp}(-i\sqrt{\alpha}\bar{t})]$, where $\sqrt{\alpha}$ is a non-dimensional frequency. Considering the above conditions, the linearized and non-dimensionalized governing equation and boundary conditions are obtained  \cite[e.g.][]{Meylan2002} with dropped overbars as
\begin{eqnarray}
\vspace{1mm}%
& \displaystyle \nabla^2\phi=0 & -h\le z\le 0, \label{eq1L} \\ %
\vspace{1mm}%
&\displaystyle \frac{\partial \phi}{\partial z}=0&  z=-h, \label{eq1B} \\
\vspace{1mm}%
&\displaystyle \frac{\partial \phi}{\partial z}-\alpha \phi=0&  z=0, r\ge b,\label{eq1F} \\ %
&\displaystyle \bigg(\beta^{(k)}\nabla_\perp^4-\alpha\gamma^{(k)}+1\bigg)\frac{\partial \phi}{\partial z}-\alpha \phi=0\quad& z=0, R^{(k+1)}\le r < R^{(k)}, \label{eq1Ck}%
\end{eqnarray}%
%
where $\beta^{(k)}=D^{(k)}/(\rho_wga^4)$ and $\gamma^{(k)}=m_c^{(k)}/(\rho_w a)$ are non-dimensional flexural rigidity and mass of $k$-th layer, respectively. 



\subsection{Spectral decomposition of velocity potential}
In order to derive a spectral method solution for the above system of equations (i.e. Eqs. \eqref{eq1L}-\eqref{eq1Ck}), we use separation of variables to express the velocity potential \cite[e.g.][]{Newman1977} as,
\begin{equation}
    \phi(r,\theta,z)=R(r)\Theta(\theta)Z(z).
    \label{eq:00-2}
\end{equation}
%
Substituting Eq. (\ref{eq:00-2}) in Eqs. \eqref{eq1L}-\eqref{eq1Ck}, the dispersion relations of water waves and elastic waves on the plate are obtained as
\begin{eqnarray}
\alpha=\left\{\begin{array}{lr}
\vspace{1mm}
\displaystyle k_0\tanh{k_0h} &n=0,\\
\vspace{1mm}
\displaystyle -k_n\tan{k_nh} & n> 0,
\end{array}\right.
\label{eq:02}
\end{eqnarray}
and
\begin{eqnarray}
\displaystyle\frac{\alpha}{\beta^{(k)}\mu_n^{(k)4}-\alpha\gamma^{(k)}+1}=\left\{\begin{array}{lr}
\vspace{1mm}
\displaystyle \mu_0^{(k)}\tanh{\mu_0^{(k)}h} &n=0,\\
\vspace{1mm}
\displaystyle -\mu_n^{(k)}\tan{\mu_n^{(k)}h} & n=-1,-2,n> 0.
\end{array}\right.
\label{eq:03}
\end{eqnarray}
Equation (\ref{eq:02}) is the dispersion relation of water waves, where $k_n$ $(n=0,1,2,\cdots)$ is the wave number of water waves.
The wave number $k_n$ in Eq. \eqref{eq:02} has infinite number of positive and real solutions, where $k_0$ denotes progressive waves, and $k_1$, $k_2\cdots$ correspond to local waves (i.e. evanescent waves). Equation (\ref{eq:03}), on the other hand, is the dispersion relation of elastic waves for the $k$-th layer of the plate, where $\mu^{(k)}_n$ is the wave number. Similarly, the wave number $\mu^{(k)}_n$ in Eq. \eqref{eq:03} has infinite number of positive and real solutions $\mu^{(k)}_0,\mu^{(k)}_1,\mu^{(k)}_2\cdots$, and additionally two complex solutions $\mu^{(k)}_{-2}$ and $\mu^{(k)}_{-1}$ where $\mu^{(k)}_{-1}=(\mu^{(k)}_{-2})^*$ and the real parts are positive.
Here, $\mu^{(k)}_0$ indicates progressive waves, $\mu^{(k)}_1,\mu^{(k)}_2\cdots$ are wave numbers of local waves, and $\mu^{(k)}_{-2}$ and $\mu^{(k)}_{-1}$ represent damped waves \cite[e.g.][]{Fox1994}. Using the above dispersion relations, the solutions for $Z(z)$ are given as
\begin{eqnarray}
Z(z)=f_n(z)=\left\{\begin{array}{lr}
\vspace{1mm}
\displaystyle \frac{\cosh{k_0(z+h)}}{\cosh k_0h}&n=0,\\
\vspace{1mm}
\displaystyle \frac{\cos{k_n(z+h)}}{\cos k_nh} &n> 0,\\
\end{array}\right.
\label{eq:04}
\end{eqnarray}
and
\begin{eqnarray}
Z(z)=F^{(k)}_n(z)=\left\{\begin{array}{lr}
\vspace{1mm}
\displaystyle \frac{\cosh{\mu^{(k)}_0(z+h)}}{\cosh\mu^{(k)}_0h}&n=0,\\
\vspace{1mm}
\displaystyle \frac{\cos{\mu^{(k)}_n(z+h)}}{\cos\mu^{(k)}_nh} &n=-1,-2,n> 0,
\end{array}\right.
\label{eq:05}
\end{eqnarray}
where $f_n(z)$ is the $z$-function of the water wave region ($r\ge b$), and $F^{(k)}_n(z)$ is the $z$-function of the $k$-th layer region $(R^{(k+1)}\le r < R^{(k)})$. 
Next, the solution for $\Theta(\theta)$ is given as $\Theta(\theta)={\rm exp}(\pm im\theta)$ for $m=0,1,\cdots$.
Lastly, solutions with respect to $r$ are given from Bessel differential equations. Since solutions are bound by the radiation condition (i.e. only progressive waves survive at far-field), the velocity potential of water wave region $\phi_w$ is given as
\begin{eqnarray}
\displaystyle \phi_{w}(r,\theta,z)=\frac{1}{i\sqrt{\alpha}}\sum_{m=-\infty}^{\infty}\bigg\{a_{m0}H_m^{(1)}(k_0r)f_0(z)+\sum_{n=1}^\infty a_{mn}K_m(k_nr)f_n(z)\bigg\}e^{im\theta},
\label{eq:06}
\end{eqnarray}
where $H_m^{(1)}(\cdot)$ is the Hankel function of the first kind, $K_m(\cdot)$ is the modified Bessel function of second kind, and $a_{mn}$ is an unknown coefficient of the velocity potential $\phi_w$. The coefficient $1/i\sqrt{\alpha}$ is used for normalizing $\phi_w$ and that of incident waves. Note that the velocity potential of the $k$-th layer region $\phi_c^{(k)}$ is 
\begin{eqnarray}
\displaystyle \phi_{c}^{(k)}(r,\theta,z)&=&\frac{1}{i\sqrt{\alpha}}\sum_{m=-\infty}^{\infty}\bigg\{b^{(k)}_{m0}J_m(\mu^{(k)}_0r)F^{(k)}_0(z)+\hspace{-4mm}\sum_{n=-2,n\ne 0}^\infty \hspace{-3mm}b^{(k)}_{mn}I_m(\mu^{(k)}_nr)F^{(k)}_n(z)\nonumber\\
&&+c^{(k)}_{m0}H_m^{(1)}(\mu^{(k)}_0r)F^{(k)}_0(z)+\hspace{-4mm}\sum_{n=-2,n\ne 0}^\infty \hspace{-3mm}c^{(k)}_{mn}K_m(\mu^{(k)}_nr)F^{(k)}_n(z)\bigg\}e^{im\theta},
\label{eq:07}
\end{eqnarray}
where $J_m(\cdot)$ is the Bessel function of first kind, $I_m(\cdot)$ is the modified Bessel function of first kind, and $b_{mn}^{(k)}$ and $c_{mn}^{(k)}$ are unknown coefficients.
Considering incident waves coming from $\theta=\pi$ as $\eta_{inc}={\rm Re}[\exp(ik_0x)\exp(-i\sqrt{\alpha}t)]$, the corresponding velocity potential $\phi_{inc}$ becomes
\begin{eqnarray}
\displaystyle \phi_{inc}(r,\theta,z)=\frac{1}{i\sqrt{\alpha}}e^{ik_0x}f_0(z)=\frac{1}{i\sqrt{\alpha}}\sum_{m=-\infty}^{\infty}i^mJ_m(k_0r)f_0(z)e^{im\theta}.
\label{eq:08}
\end{eqnarray}
In summary, the solution $\phi$ to this problem, using the spectral method  \cite[e.g.][]{Peter2004}, is given as
\begin{eqnarray}
\phi(r,\theta,z)=\left\{\begin{array}{lr}
\vspace{1mm}
\displaystyle \phi_{inc}+\phi_w&r\ge b,\\
\vspace{1mm}
\displaystyle \phi_{c}^{(k)}&R^{(k+1)}\le r < R^{(k)}.\\
\end{array}\right.
\label{eq:09}
\end{eqnarray}
The unknown coefficients $a_{mn}$, $b_{mn}^{(k)}$ and $c_{mn}^{(k)}$ are numerically determined to satisfy further boundary conditions. A numerical approach for solving this problem will be discussed in the numerical approach section (\S\ref{section:numerical-approach}).


\subsection{Energy of scattered-waves at far-field}
Performance of the plate as a cloaking device is quantified by the energy of scattered-waves \cite[]{Iida2014}.
To obtain the energy of scattered-waves, we consider a control surface $S_\infty$ at the far-field that surrounds the cylinder and the plate and then calculate the flow of energy passing through this surface.
The non-dimensional energy $W$ is obtained \cite[]{Maruo1960, Kashiwagi2005} as
\begin{eqnarray}
\displaystyle W=-\iint_{S_\infty}\overline{\frac{\partial \Phi}{\partial t}\frac{\partial \Phi}{\partial n}}dS
\simeq -\int_{-h}^0dz \int_0^{2\pi}\overline{\frac{\partial \Phi}{\partial t}\frac{\partial \Phi}{\partial r}}rd\theta.
\label{eq:10}
\end{eqnarray}
Since the control surface is far from the cylinder and the plate, local waves created by structures and the waves are fully attenuated, and as a result, the velocity potential at far-field $\Phi_{far}$ becomes
\begin{eqnarray}
\displaystyle   \Phi_{far}(r,\theta,z,t) = {\rm Re} \left[  \phi_{far}(r,\theta,z)e^{-i  \sqrt{\a} t}  \right], 
\label{eq:11}
\end{eqnarray}
where
\begin{eqnarray}
\displaystyle   \phi_{far}(r,\theta,z) = \f{1}{i \sqrt{\a}} \sum_{m=-\infty}^{\infty} \bigg\{  i^mJ_m(k_0r)+a_{m0}H_m^{(1)}(k_0r)\bigg\}f_0(z) 
   e^{im\theta}.
\label{eq:12}
\end{eqnarray}
Substituting Eq. (\ref{eq:11}) in Eq. (\ref{eq:10}), we find the energy 
\begin{eqnarray}
\displaystyle W&=&-\frac{ik_0}{8C_0\sqrt{\alpha}}\int_0^{2\pi} \sum_{p=-\infty}^\infty\sum_{q=-\infty}^\infty
\bigg\{\bigg( (-i)^qJ_q+a_{q0}^*H_q^{(2)}\bigg)\bigg(i^pJ_p'+a_{p0} H_p^{(1)'}\bigg) \nonumber\\
\displaystyle &&-\bigg(i^pJ_p+a_{p0}H_p^{(1)}\bigg)\bigg((-i)^qJ_q'+a_{q0}H_q^{(2)'} \bigg)
 e^{ip\theta}e^{-iq\theta}\bigg\}rd\theta,
\label{eq:13}
\end{eqnarray}
where 
\begin{eqnarray}
\displaystyle C_0=\frac{k_0^2}{\alpha+(k_0^2-\alpha^2)h}.
\label{eq:14}
\end{eqnarray}
Note that the orthogonal relations and  Wronskian formulae are \cite[see][]{Kashiwagi2001} 
\begin{eqnarray}
\displaystyle &&\int_0^{2\pi}e^{ip\theta}e^{-iq\theta}d\theta=2\pi \delta _{pq},
\label{eq:15}\\
\displaystyle &&\left\{
  \begin{array}{l}
  \vspace{2mm}
\displaystyle J_mH_m^{(1)'}-J_m'H_m^{(1)}=\frac{2i}{\pi k_0r},\\
\vspace{2mm}
\displaystyle J_m'H_m^{(2)}-J_mH_m^{(2)'}=\frac{2i}{\pi k_0r},\\
\displaystyle H_m^{(1)'}H_m^{(2)}-H_m^{(1)}H_m^{(2)'}=\frac{4i}{\pi k_0r},
  \end{array} 
\right. 
\label{eq:16}
\end{eqnarray}
where $\delta_{pq}$ is the Kronecker delta.
Using Eqs. (\ref{eq:15}) and (\ref{eq:16}), we can simplify the energy term to
\begin{eqnarray}
\displaystyle W=\frac{1}{C_0\sqrt{\alpha}}\sum_{m=-\infty}^\infty\bigg\{\text{Re} [i^m a_{m0}^*]+|a_{m0}|^2 \bigg\}=0.
\label{eq:17}
\end{eqnarray}
The above equation is the energy conservation principle, and thus it must be zero (we later use this fact to check the numerical accuracy of our code). Note that since $a_{m0}$ is the amplitude of scattered-waves, the second term in Eq. (\ref{eq:17}) indicates the energy transferred to scattered-waves. As a result, the energy of scattered-waves $W_s$ is given as 
\begin{eqnarray}
\displaystyle W_s=\frac{1}{C_0\sqrt{\alpha}}\sum_{m=-\infty}^\infty|a_{m0}|^2.
\label{eq:18}
\end{eqnarray}
The energy of scattered-waves obtained here (Eq. \eqref{eq:18}) is equal to the form of the Kochin function in \cite{Newman2014}. We define the cloaking factor $\mathcal{F}_{clk}$  by the ratio of the energy of scattered-waves of the cylinder itself $W_{cyl}$ and that of the cylinder with the cloak $W_{clk}$  \cite[]{Porter2014}, i.e. 
\begin{eqnarray}
\displaystyle \mathcal{F}_{clk}=\frac{W_{clk}}{W_{cyl}}.
\label{eq:211} 
\end{eqnarray}
If  $\mathcal{F}_{clk}<1$ then the energy scattered by the cylinder is decreased by the cloak and clearly, the perfect cloaking is achieved as $\mathcal{F}_{clk}=0$.

Next, we calculate the wave drift force acting on the cylinder. The wave drift force is the time-averaged second-order hydrodynamic force calculated by the first-order velocity potential.
Considering the control surface at far-field, the wave drift force can be calculated by the momentum conservation principle \cite[]{Maruo1960}  
\begin{eqnarray}
\displaystyle \overline{F_x}=\overline{\iint_{S_H}pn_xdS}=-\overline{\iint_{S_\infty}\bigg(pn_x+\frac{\partial \Phi}{\partial x}u_n\bigg)dS},
\label{eq:19}
\end{eqnarray}
where $\overline{F_x}$ is the non-dimensional wave drift force, $S_H$ is the surface of the body, $p$ is pressure, $n_x$ is the $x$-component of normal vector, and $u_n$ is the velocity in the direction of normal vector.
Using similar analytical expansion to the energy of scattered-waves, the formula for the wave drift force is obtained \cite[]{Kashiwagi2001} 
\begin{eqnarray}
\displaystyle \overline{F_x}\simeq \frac{k_0}{2C_0\alpha}\sum_{m=-\infty}^{\infty}\text{Im}[2a_{m0}a^*_{m+1,0}+i^ma^*_{m+1,0}+(-i)^{m+1}a_{m0}].
\label{eq:20}
\end{eqnarray}
Since the wave drift force is calculated by the amplitude of scattered-waves $a_{m0}$, the wave drift force acting on bodies becomes very small when there is no scattered-wave.

\section{Numerical approach}
\label{section:numerical-approach}
The spectral solution (i.e. Eqs. \eqref{eq:06} and \eqref{eq:07}) to this problem (Eqs. \eqref{eq1L}-\eqref{eq1Ck}) has unknown coefficients (i.e. $a_{mn}$, $b_{mn}^{(k)}$ and $c_{mn}^{(k)}$) that can be found by satisfying the boundary conditions.
At first, we approximate the solution by truncating the infinite number of modes in Eq. (\ref{eq:09}) to orders $M$ and $N$ for azimuthal and radial terms respectively. 
The numbers $M$ and $N$ are decided such that the solution is converged and the results do not change by increasing $M$ and $N$ any further. For each azimuthal mode, we have $N+1$ unknowns in the water wave region (i.e. $a_{mn}$ for $n=0,1,2,\cdots,N$), and $2K(N+3)$ unknowns in the plate region (i.e. $b^{(k)}_{mn}$ and $c^{(k)}_{mn}$ for $k=1,2,\cdots,K$ and $n=-2,-1,0,1,\cdots,N$).
In total, the number of unknown coefficients is $N+1+2K(N+3)$, and thus the same number of boundary conditions are required for solving them.

We utilize an eigenvalue matching method \cite[]{Peter2004} to determine the unknown coefficients. The form of the velocity potential in Eq. (\ref{eq:09}) depends on the radial direction $r$, and matching of each quantity at the boundaries should be considered. Since these matching boundary conditions are valid throughout the depth, $z$-function $f_\ell(z)$ $(\ell=0,1,\cdots,N)$ is selected as a set of basis functions. These functions are multiplied by the velocity potential, and these are integrated throughout the water depth. Resultant functions with respect to $z$ are defined as 
\begin{eqnarray}
\displaystyle &&A_{n\ell} \equiv \int_{-h}^{0} f_n(z) f_\ell(z) \d z =\frac{1}{2}\bigg( \frac{\cos{k_nh}\sin{k_nh}+k_nh}{k_n\cos^2{k_nh}}\bigg)\delta_{n\ell},
\label{eq:21}\\
\displaystyle &&B_{n\ell}^{(k)} \equiv \int_{-h}^{0} F_n^{(k)}(z)f_\ell(z) \d z= \frac{k_\ell\sin{k_\ell h}\cos{\mu_n^{(k)}h}-\mu_n^{(k)}\cos{k_\ell h}\sin{\mu_n^{(k)}h} }{(k_\ell^2-\mu_n^{(k)2})\cos{k_\ell h}\cos{\mu_n^{(k)}h}}.
\label{eq:22}
\end{eqnarray}
Note that wave numbers in Eqs (\ref{eq:21}) and (\ref{eq:22}) must be replaced as $k_0\to ik_0$ and $\mu_0^{(k)}\to i\mu_0^{(k)}$ when $n=0$.
Using Eqs (\ref{eq:21}) and (\ref{eq:22}) for each boundary condition, the number of conditions becomes $N+1$. Next, we need to satisfy the following boundary conditions:
\begin{enumerate}
  \item 
Matching conditions of the velocity potential between the free surface and the plate ($N+1$ equations).
  \item 
Matching conditions of the radial derivative of the velocity potential between the free surface and the plate ($N+1$ equations).
  \item 
Matching conditions of the velocity potential between adjacent layers ($(K-1)(N+1)$ equations).
  \item 
Matching conditions of the radial derivative of the velocity potential between adjacent layers ($(K-1)(N+1)$ equations).
  \item 
No flux conditions at the surface of the cylinder ($N+1$ equations).
  \item 
Matching conditions of the wave elevation between adjacent layers ($K-1$ equations).
  \item 
Matching conditions of the radial derivative of the wave elevation between adjacent layers ($K-1$ equations).
  \item 
Matching conditions of the bending moment between adjacent layers ($K-1$ equations).
  \item 
Matching conditions of the shear force between adjacent layers ($K-1$ equations).
  \item 
Free-free beam conditions; zero bending moment and shear force (4 equations).
\end{enumerate}
Matching the above list of boundary conditions, we obtain $N+1+2K(N+3)$ equations which is the same as the number of unknowns. Solving these equations, all unknown coefficients are found. Details of these boundary conditions are shown in Appendix A. For completeness, here, we calculate the bending moment and equivalent shear force on the plate using the surface elevation of the plate. These quantities for the radial direction acting on the $k$-th layer are given as
\begin{eqnarray}
  M^{(k)}_r (r, \theta, t)  &=& \text{Re}\bigg[\sum_{m=-\infty}^{\infty} \mathcal{M}_m^{(k)}(r) e^{im\th}e^{-i\sqrt{\alpha} t} \bigg],\label{eq23:01}\\
  V^{(k)}_r (r, \theta, t) & =& \text{Re}\bigg[\sum_{m=-\infty}^{\infty} \mathcal{V}_m^{(k)}(r) e^{im\th} e^{-i\sqrt{\alpha} t} \bigg],\label{eq23:02} 
\end{eqnarray}
where $\mathcal{M}_m^{(k)} (r)$ and $\mathcal{V}_m^{(k)} (r)$ are written as
\begin{eqnarray}
\displaystyle&& \mathcal{M}_m^{(k)} (r) = -\left[ \nabla^2_\perp -\f{1-\nu}{r}\lp \f{\p}{\p r} -\f{m^2}{r} \rp \right] \psi^{(k)}(r),\label{eq23:03} \\
\displaystyle&&\mathcal{V}_m^{(k)}(r) =-\left[ \f{\p}{\p r}\nabla^2_\perp -m^2\f{1-\nu}{r^2}\lp \f{\p}{\p r} - \f{1}{r} \rp \right] \psi^{(k)}(r).
\label{eq23:04} 
\end{eqnarray}
In the above equations, note that 
\begin{eqnarray}
\displaystyle \psi^{(k)}(r)=&&b^{(k)}_{m0}G^{(k)}_0J_m(\mu^{(k)}_0r)+\hspace{-4mm}\sum_{n=-2,n\ne 0}^\infty \hspace{-3mm}b^{(k)}_{mn}G^{(k)}_nI_m(\mu^{(k)}_nr)\nonumber\\
&&+c^{(k)}_{m0}G^{(k)}_0H_m^{(1)}(\mu^{(k)}_0r)+\hspace{-4mm}\sum_{n=-2,n\ne 0}^\infty \hspace{-3mm}c^{(k)}_{mn}G^{(k)}_nK_m(\mu^{(k)}_nr),\label{eq23:05} 
\end{eqnarray}
where
\begin{eqnarray}
\displaystyle G_n^{(k)}=\frac{\beta^{(k)}}{\beta^{(k)}\mu_n^{(k)4}-\alpha \gamma^{(k)}+1}.\label{eq23:06}
\end{eqnarray}

\section{Evolutionary optimization of the plate}
\label{section:evolutionary}
Our goal here is to cancel the energy of scattered-waves from the cylinder by optimizing the buoyant plate parameters. In other words, we optimize the parameters of the plate to minimize the energy of scattered-waves through a meta-heuristic approach. We use an evolutionary optimization method, specifically, the real-coded genetic algorithm (RGA) based on the unimodal normal distribution crossover and minimal generation gap \cite[]{Ono1999}. We rewrite the problem here as
\begin{eqnarray}
\begin{array}{llr}
{\rm minimize}&W_{clk}, &\\
{\rm subject\; to}&0.01\le \beta^{(k)}\le 0.5&k=1,2,\cdots,K,\\
&0.01\le \gamma^{(k)}\le 0.5&k=1,2,\cdots,K,
\end{array}
\label{eq4:01}
\end{eqnarray}
where optimum non-dimensional flexural rigidity $\beta^{(k)}$ and mass $\gamma^{(k)}$ are sought under selected numerical conditions: the wave number $k_0$, the water depth $h$, the outermost radius of the plate $b$, the number of layers $K$, and the Poisson's ratio of the plate $\nu$. Note that the above minimization is done at a certain wave number of incident waves. In order to show the full frequency response of the structures, we later find the response of our structures at wave numbers different from the optimized one.

\begin{figure}
 \includegraphics[width=5.0in]{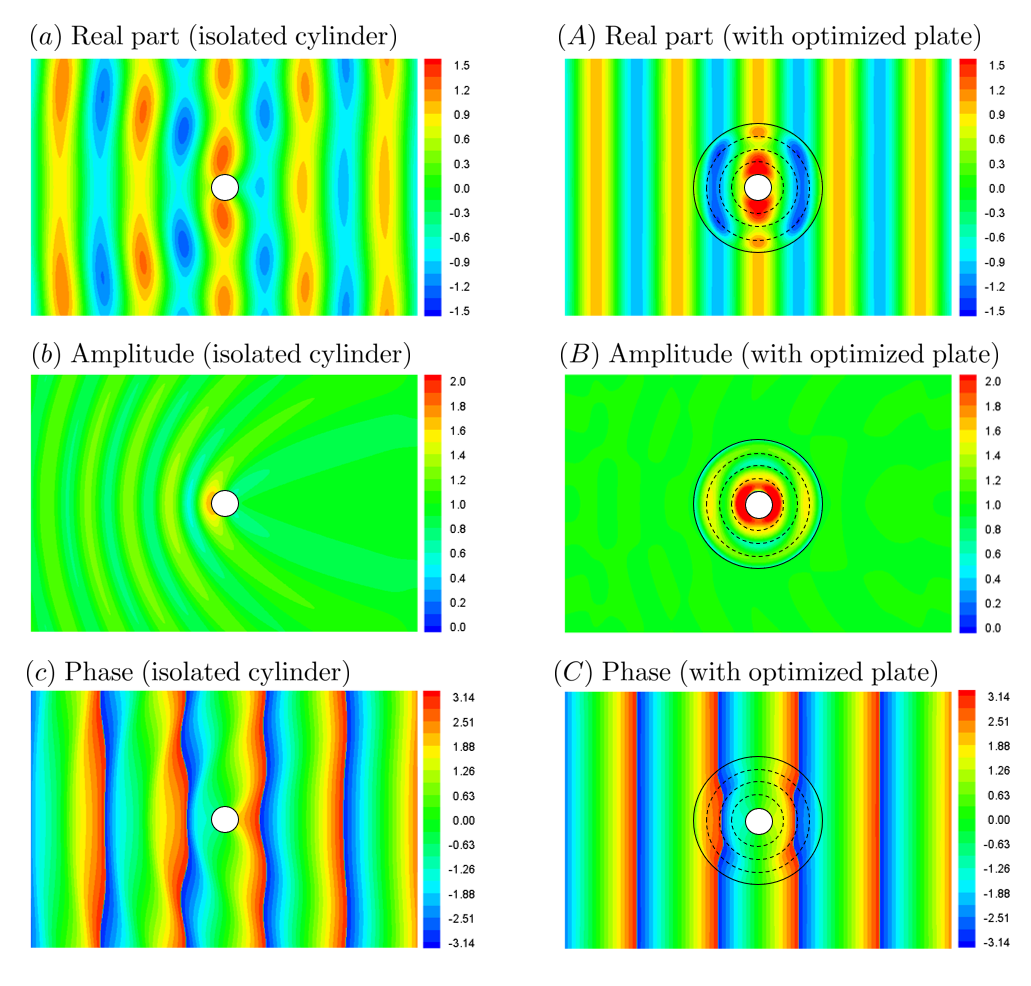}
 \caption{The real part (a,A), amplitude (b,B), and phase field (c,C) of the wave solution for an isolated cylinder (left column) and the cylinder surrounded by the optimized plate (right column). The cloaking plate consists of $K=4$ layers where the flexural rigidity and mass  of each layer is optimized  to minimize the scattering energy for the wave number $k_0=1.0$, i.e. case I.  }
 \label{fig2}
\end{figure}

\section{Results and discussions}

To show the effectiveness of the proposed multi-layered cloak, numerical simulations are carried out. In order to consider the finite depth and deep water waves, we assume $h/\lambda =1.0$,  where  $h$ is the water depth and  $\lambda$ is the wavelength. We fix the Poisson's ratio of the plate $\nu=0.25$, and also consider the same radial width for each layer of the plate, i.e.  $R^{(k)}-R^{(k+1)} = \text{const.}, \forall k=1,\cdots, K$, where $K$ is the total number of the plate's layers.
Here, we aim to cloak the cylinder at the wave number $k_0=1.0$.
The energy of scattered-waves of the isolated cylinder at this wave number is $W_{cyl}=0.500$.
To suppress this energy, we optimize plate parameters, i.e. flexural rigidity $\beta^{(k)}$, and mass $\gamma^{(k)}$, using the evolutionary strategy at (\S\ref{section:evolutionary}). In order to optimize the plate parameters, we consider three different cases: (1) optimizing both flexural rigidity $\beta^{(k)}$ and mass $\gamma^{(k)}$ for all of the layers of the plate $(k=1,2,\cdots,K)$. We call this case I; (2) optimizing the flexural rigidity for each layer of the plate ($\beta^{(k)}$, $k=1,2,\cdots,K$) while keeping the mass as a constant for all layers (i.e. $\gamma^{(k)}=\gamma^{(1)},$ $k=2,3,\cdots,K$). We denote this as case II; and lastly (3) optimizing mass of the plate for each layer of the plate ($\gamma^{(k)}$, $k=1,2,\cdots,K$) while keeping the flexural rigidity as a  constant for all layers ($\beta^{(k)}=\beta^{(1)},$ $k=2,3,\cdots,K$). We call this case III. 

\begin{figure}
 \centering
 \includegraphics[width=5.4in]{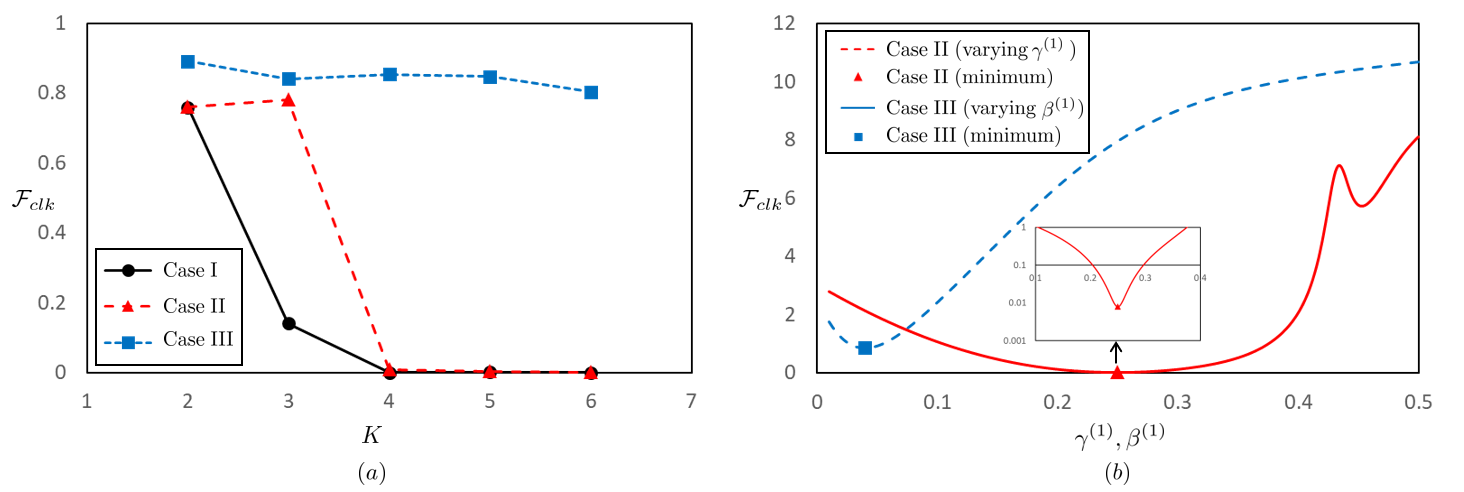}
 \caption{(a) Comparison of cloaking factor $\mathcal{F}_{clk}$ versus the number of layers $K$ for the case I, when both flexural rigidity and mass are optimized for each layer of the plate;
case II when the flexural rigidity is optimized for each layer while the mass stays as a constant for all layers; and 
case III, when the mass is optimized on every layer while the flexural rigidity remains a constant for all layers.
(b) Sensitivity study of cloaking factor to flexural rigidity and mass for plates with $K=4$ layers for case II with varying mass $\gamma^{(1)}$ and case III with varying flexural rigidity $\beta^{(1)}$. The inset figure shows the cloaking factor around the minimum value of case II in a semi-log plot. The optimized solutions found for $K=4$ in the left are highlighted by the square and triangle in the right figure.
}
\label{fig3}
\end{figure}
We visualize the wave field around an isolated cylinder and also a cylinder with the cloaking plate surrounding it as Fig. \ref{fig2}. The optimized plate shown in Fig. \ref{fig2} consists of $K=4$ layers with the outermost radius $b=5.0$, where both flexural rigidity and the mass of the plate is optimized to minimize the energy of scattered-waves (case I). The optimized plate yields $W_{clk}=0.004$, with a cloaking factor of $\mathcal{F}_{clk}=0.008$. The comparisons of wave patterns between the isolated cylinder (left column) and the cylinder with the plate (right column) are shown.  The real parts of the wave elevation around the cylinder are shown in Figs. \ref{fig2}(a) and (A), the complex amplitudes of the wave elevation are shown in Figs. \ref{fig2}(b) and (B), and the phases of wave elevation are shown in Figs. \ref{fig2}(c) and (C). Contrary to the isolated cylinder which generates outgoing scattered-waves, the cylinder with a surrounding plate has no visually identifiable outgoing wave in the wave amplitudes field around it (Figs. \ref{fig2}(a) and (A)). Similarly, we observe that the isolated cylinder modulates the wave phase field, however, the cylinder with the plate has a phase-field that matches that of incident waves.
Since the wave amplitude and phase at the plate are almost symmetric with respect to the $y$-axis, it yields an almost zero wave drift force (we will quantify the forces exerted on the cylinder later in this section). As a result, the optimized multi-layered buoyant plate is significantly reducing the energy of scattered-waves of deep water waves and cloaking the cylinder from the incident wave. It is noted that since the cloaking plate here is 
axisymmetric, its functionality is independent of incoming waves' direction and as a result, is omnidirectional.


\begin{figure}
 \centering
 \includegraphics[width=3.6in]{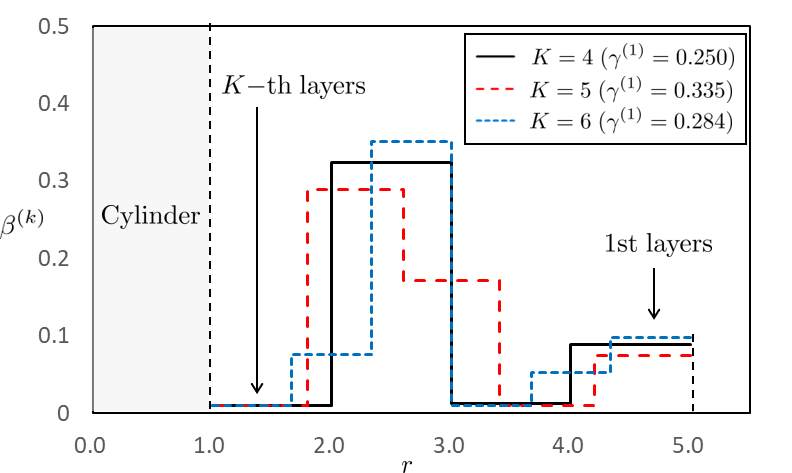}
 \caption{Spatial distributions of flexural rigidities $\beta^{(k)}$ for the cloaking plate. The cloaking plate extends at $1.0\le r\le 5.0$ where the cylinder is inside $r\leq 1.0$. We assume a constant mass density across the layers, and variable flexural rigidity (case II). We optimize the cloaking plate parameters for different number of layers  $K=4,5,$ and $6$. The value found for the optimized $\gamma$ is noted in the legend, and the flexural rigidity profile is plotted across the layers.}
 \label{fig4}
\end{figure}

Next, we investigate the effect of the number of layers $K$ on the effectiveness of the cloaking plate. We fix the outermost radius of the plate at $b=5.0$ and we plot the cloaking factor versus the number of layers for the optimized plate for the three cases (I, II, \& III) in Fig. \ref{fig3}(a). Interestingly, when the layer number is $K\ge 4$, the cloaking factors in cases I and II become less than 0.01. However, the cloaking factor in case III does not decrease significantly and the cloaking factor remains above 0.8. As a result, varying flexural rigidity $\gamma^{(k)}$ is essential for manipulating waves and creating a cloaking buoyant plate. Since the case I has a  cloaking factor smaller than case II, optimizing the mass $\beta^{(k)}$ helps to achieve a better cloak, nevertheless only optimizing the mass is insufficient to realize an effective cloaking plate. To investigate the sensitivity of the cloaking factor to the flexural rigidity and mass, we calculate the cloaking factor while modulating these parameters. 
The sensitivity of the factor to these parameters is shown in Fig. \ref{fig3}(b).
The result of case II against varying mass $\gamma^{(1)}$, and that of case III against varying flexural rigidity $\beta^{(1)}$ are plotted; optimum values are marked by square and triangle respectively. 
The inset figure shows the semi-log plot of the cloaking factor around the minimum value for case II.
According to the semi-log graph, the cloaking factor for case II increases as the mass $\gamma^{(1)}$ slightly changes from the optimum value.
When the mass $\gamma^{(1)}$ and flexural rigidity $\beta^{(1)}$ are far from the optimum values, the cloaking factors are bigger than 1.0 meaning that the structure scatters waves with an energy larger than that of the isolated cylinder. It can be concluded that the cloaking factor is sensitive to both flexural rigidity and mass of the plate and large deviation can be detrimental to the efficacy of the cloaking plate. 


The spatial distribution of the flexural rigidity in case II is shown in Fig. \ref{fig4} where it shows the structural flexural rigidity profile of the multi-layered plate. Note that $r\le 1.0$ is the cylinder region; the plate is at $1.0\le r\le 5.0$. Results of $K=4,5$ and $6$ are plotted and fix values of mass are noted in the legend. For $K=4$, the plate has two sets of peaks and valleys with the first peak being larger than the second one.
The values of valleys are almost at $\beta = 0.01$ which is the lower limit of the physical parameter (see Eq. (\ref{eq4:01})).
Interestingly, the rigidity profile follows a similar trend for different number of layers. The two sets of peaks and valleys are not achievable with only three number of layers, and this might be the reason why cloaking can not be achieved for $K<4$ (see Fig. \ref{fig3}(a)).

\begin{figure}
 \centering
 \includegraphics[width=3.6in]{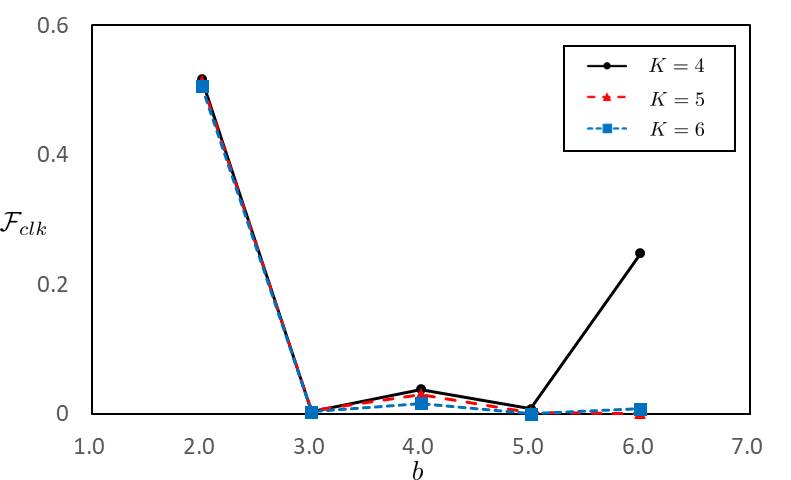}
  \caption{Cloaking factor $\mathcal{F}_{clk}$ against sizes of outermost radius $b=2.0,3.0,4.0,5.0$ and 6.0. Case II (mass $\gamma^{(k)}$ is constant as $\gamma^{(1)}$ for all layers) is considered. The influence of outermost radius size on cloaking factor is investigated using different layer numbers $K=4,5$ and 6. Note that zero cloaking factor $\mathcal{F}_{clk}=0$ yields perfect cloaking.}
\label{fig5}
\end{figure}
In addition to the number of layers, we analyze the influence of the outermost radius $b$ on the cloaking factor. The cloaking factor for case II against the outermost radius $b$ is shown in  Fig. \ref{fig5}.
When the outermost radius is $b=2.0$, the factor is not fully reduced even with increasing the number of layers $K$. The cloaking factor for $K=4$ and $b=6.0$ is also not small enough, while results of $K=5$ and $6$ are less than 0.01. When $b=4.0$, the cloaking factors for different values of $K$ are $\mathcal{F}_{clk}|_{K=4}=0.039$, $\mathcal{F}_{clk}|_{K=5}=0.030$, and $\mathcal{F}_{clk}|_{K=6}=0.017$; the use of the larger layer number helps us to achieve a smaller cloaking factor. In summary, it is observed that the size of the plate $b$ affects the cloaking factor and cloaking can not be realized using a small plate size. Additionally, using a bigger plate size does not necessarily indicate a better cloaking result. An optimized plate size exists that should be selected for efficient cloaking.

\begin{figure}
 \centering
 \includegraphics[width=5.4in]{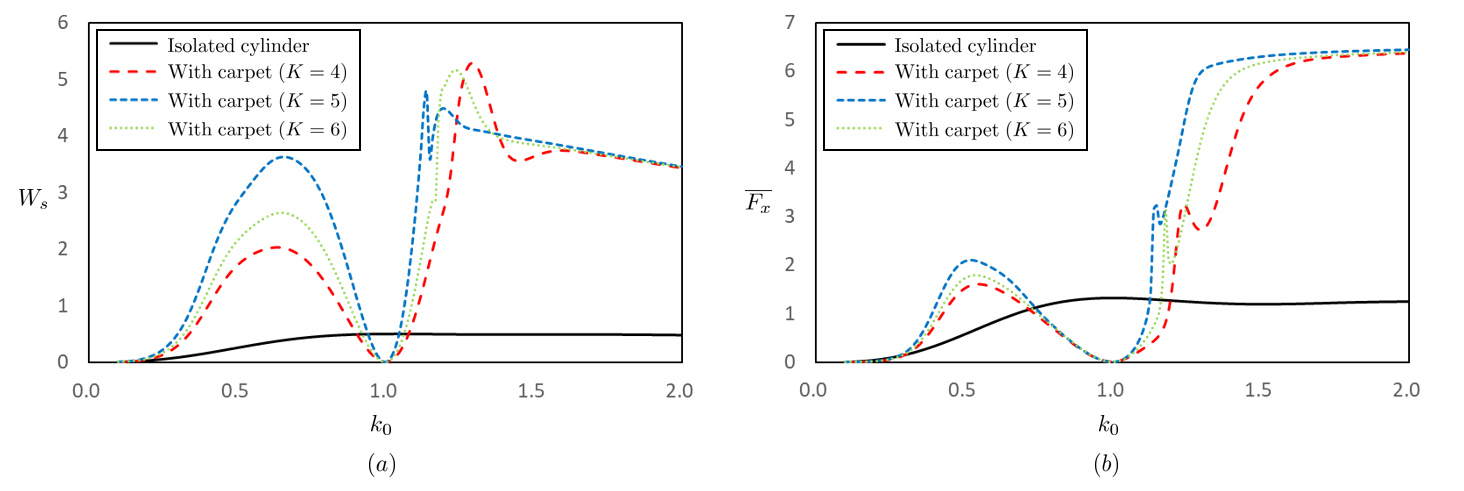}
 \caption{(a) Energy of scattered-waves $W_s$ and (b) wave drift force in $x$-direction $\overline{F_x}$ against wave number $k_0$. Results are compared among the isolated cylinder and the cylinder with the surrounding cloaking plate. Here, the outermost radius of the plate is $b=5.0$.  We consider case II (mass $\gamma^{(k)}$ being constant across layers while $\gamma^{(1)}$ can vary) with the number of layers $K=4,5$ and 6. plate parameters are optimized to minimize energy of scattered-waves at wave number $k_0=1.0$. }
 \label{fig6}
\end{figure}

Lastly, we analyze the frequency response of the cloaking plate, i.e. the behavior of the structures for wave numbers or wave frequencies other than the value that it is optimized for.
The energy of scattered-waves is shown in Fig. \ref{fig6}(a), and; the wave drift force acting in $x$-direction is shown in Fig. \ref{fig6}(b). The result of the isolated cylinder and the results of the cylinder with the multi-layered cloak are compared. The outermost radius of the plate is fixed at $b=5.0$, and physical parameters in case II are used. 
Looking at $k_0=1.0$, the wave drift force of the isolated cylinder is $\overline{F_x}|_{cyl}=1.330$ and the cloaked wave drift forces by different $K$ are $\overline{F_x}|_{K=4}=0.010$, $\overline{F_x}|_{K=5}=0.004$ and $\overline{F_x}|_{K=6}=0.001$; 99.9\% reduction of wave drift force is achieved at most.
Therefore, both energy of scattered-waves and wave drift force are dramatically reduced and become almost zero using the multi-layered cloak at the target wave number $k_0=1.0$. Note that the wave drift force is the second-order force based on the law of action and reaction of wave scattering. As a result, the wave drift force is not acting on the structures if no scattered-wave is generated, or if the scattered-wave field is $y$-symmetric as shown in Fig. \ref{fig2}(A).  As for frequency responses, the drift force of the cloaked cylinder is smaller than that of the isolated cylinder around the target wave number with a bandwidth of $\Delta k=0.5$ (in case of $K=4$). However, these values are larger than those of the isolated cylinder outside this band, meaning that energy scattered by the structures is larger than that of the isolated cylinder.
Note that the energy of scattered-waves and the wave drift force converge as  $k_0\to 0$ and $k_0\to 2.0$. Interestingly, using a larger number of layers $K$ does not indicate a smaller wave drift force for the frequency bands. For realizing an optimized cloak in a frequency band, a multi-objective optimization might be implemented instead of our one frequency optimization. Since we can increase the degrees of freedom for controlling wave propagation by easily increasing the number of layers in the multi-layered plate, such plate design can potentially achieve a broadband cloak from deep water waves.

\section{Conclusion}
We present cloaking of a bottom-mounted cylinder from water waves in deep-sea using a multi-layered elastic plate floating on the surface around the cylinder.
In the governing equation of surface gravity waves, the sea bed topography and the gravitational acceleration are the only physical parameters controlling the trajectory of wave propagation; nevertheless, the effect of sea depth exponentially decreases as the depth increases, and the gravitational acceleration is physical constant. As a consequence, cloaking the offshore structure from deep water waves becomes a challenge. Here, we propose the use of a multi-layered buoyant plate to provide extra adjustable degrees of freedom for controlling wave propagation.
The plate consists of $K$ concentric annular layers with isotropic and homogeneous physical parameters.
This design enables easier experimental implementation.
The multi-layered cloak is created based on the scattering cancellation method; the plate cancels out the scattered-waves from the cylinder.
Physical parameters of each layer are optimized using an evolutionary strategy method, specifically a real-coded genetic algorithm.
A numerical calculation scheme is developed using pseudo-spectral and eigenvalue matching methods, and the effectiveness of the multi-layered cloak is evaluated. The plate is designed to cloak the cylinder at the specific wave number. We vary different parameters of the buoyant cloaking plate and analyzed their effects on the cloaking factor. We show that an optimum cloaking size and the number of layers exist for maximally increasing the efficiency of the cloak. We further address the sensitivity of the cloak to changes in the wave frequency and show an optimum working bandwidth exists. Since the reduction of the wave drift force is important for practical applications, we show that our proposed cloak has 99.9\% reduction of wave drift force at most. We believe that our proposed cloak has potential real-world applications in the fields of offshore industries to protect offshore structures.

\bibliographystyle{jfm}
\bibliography{ref}

\appendix
\section{Details of boundary conditions}
In the following we list the details of all boundary conditions that are used:
\begin{itemize}
  \item 
Matching conditions of the velocity potential between the free surface and the structure ($N+1$ equations):
\begin{eqnarray}
\begin{array}{l}
\displaystyle i^mJ_m(k_0b)A_{0\ell}+a_{m0}H_m^{(1)}(k_0b)A_{0\ell}+ \sum_{n=1}^{N} a_{mn} K_m(k_n b)A_{n\ell} = \\
\displaystyle \quad b_{m0}^{(1)}J_m(\mu^{(1)}_0 b)B_{0\ell}^{(1)}+ \hspace{-4mm} \sum_{n=-2,n\ne 0}^{N}\hspace{-3mm}b^{(1)}_{mn} I_m(\mu^{(1)}_n b) B_{n\ell}^{(1)} \\
\displaystyle \quad+c_{m0}^{(1)}H_m^{(1)}(\mu^{(1)}_0 b)B_{0\ell}^{(1)} + \hspace{-4mm} \sum_{n=-2,n\ne 0}^{N}\hspace{-3mm} c^{(1)}_{mn} K_m(\mu^{(1)}_n b) B_{n\ell}^{(1)}.
\end{array}
\end{eqnarray}

  \item 
Matching conditions of the radial derivative of the velocity potential between the free surface and the structure ($N+1$ equations):

\begin{eqnarray}
\begin{array}{l}
\displaystyle i^mk_0J'_m(k_0b)A_{0\ell}+a_{m0}k_0H_m^{(1)'}(k_0b)A_{0\ell}+ \sum_{n=1}^{N} a_{mn}k_n K'_m(k_n b)A_{n\ell} =  \\
\displaystyle \quad b_{m0}^{(1)}\mu^{(1)}_0J'_m(\mu^{(1)}_0 b)B_{0\ell}^{(1)}+ \hspace{-4mm} \sum_{n=-2,n\ne 0}^{N}\hspace{-3mm} b^{(1)}_{mn}\mu^{(1)}_n I'_m(\mu^{(1)}_n b) B_{n\ell}^{(1)} \\
\displaystyle \quad  +c_{m0}^{(1)}\mu^{(1)}_0H_m^{(1)'}(\mu^{(1)}_0 b)B_{0\ell}^{(1)} + \hspace{-4mm} \sum_{n=-2,n\ne 0}^{N}\hspace{-3mm}c^{(1)}_{mn}\mu^{(1)}_n K'_m(\mu^{(1)}_n b) B_{n\ell}^{(1)}.
\end{array}
\end{eqnarray}

  \item 
Matching conditions of the velocity potential between adjacent layers ($(K-1)(N+1)$ equations):

\begin{eqnarray}
\begin{array}{l}
\displaystyle b_{m0}^{(k)}J_m(\mu^{(k)}_0 R^{(k+1)})B_{0\ell}^{(k)}+ \hspace{-4mm} \sum_{n=-2,n\ne 0}^{N}\hspace{-3mm} b^{(k)}_{mn} I_m(\mu^{(k)}_n R^{(k+1)}) B_{n\ell}^{(k)} \\
\displaystyle +c_{m0}^{(k)}H_m^{(1)}(\mu^{(k)}_0 R^{(k+1)})B_{0\ell}^{(k)} +\hspace{-4mm} \sum_{n=-2,n\ne 0}^{N}\hspace{-3mm} c^{(k)}_{mn} K_m(\mu^{(k)}_n R^{(k+1)}) B_{n\ell}^{(k)}=\\
\displaystyle \quad b_{m0}^{(k+1)}J_m(\mu^{(k+1)}_0 R^{(k+1)})B_{0\ell}^{(k+1)}+ \hspace{-4mm} \sum_{n=-2,n\ne 0}^{N}\hspace{-3mm} b^{(k+1)}_{mn} I_m(\mu^{(k+1)}_n R^{(k+1)}) B_{n\ell}^{(k+1)} \\
\displaystyle \quad +c_{m0}^{(k+1)}H_m^{(1)}(\mu^{(k+1)}_0 R^{(k+1)})B_{0\ell}^{(k+1)} +\hspace{-4mm} \sum_{n=-2,n\ne 0}^{N}\hspace{-3mm} c^{(k+1)}_{mn} K_m(\mu^{(k+1)}_n R^{(k+1)}) B_{n\ell}^{(k+1)}.
\end{array}
\end{eqnarray}

  \item 
Matching conditions of the radial derivative of the velocity potential between adjacent layers ($(K-1)(N+1)$ equations):

\begin{eqnarray}
\begin{array}{l}
\displaystyle b_{m0}^{(k)}\mu^{(k)}_0J'_m(\mu^{(k)}_0 R^{(k+1)})B_{0\ell}^{(k)}+ \hspace{-4mm} \sum_{n=-2,n\ne 0}^{\infty}\hspace{-3mm}b^{(k)}_{mn}\mu^{(k)}_n I'_m(\mu^{(k)}_n R^{(k+1)}) B_{n\ell}^{(k)} \\
\displaystyle +c_{m0}^{(k)}\mu^{(k)}_0H_m^{(1)'}(\mu^{(k)}_0 R^{(k+1)})B_{0\ell}^{(k)} + \hspace{-4mm} \sum_{n=-2,n\ne 0}^{N}\hspace{-3mm}c^{(k)}_{mn}\mu^{(k)}_n K'_m(\mu^{(k)}_n R^{(k+1)}) B_{n\ell}^{(k)}=\\
\displaystyle \quad  b_{m0}^{(k+1)}\mu^{(k+1)}_0J'_m(\mu^{(i+1)}_0 R^{(k+1)})B_{0\ell}^{(k+1)}+\hspace{-4mm} \sum_{n=-2,n\ne 0}^{N}\hspace{-3mm} b^{(k+1)}_{mn}\mu^{(k+1)}_n I'_m(\mu^{(k+1)}_n R^{(k+1)}) B_{n\ell}^{(k+1)} \\
\displaystyle \quad +c_{m0}^{(k+1)}\mu^{(k+1)}_0H_m^{(1)'}(\mu^{(k+1)}_0 R^{(k+1)})B_{0\ell}^{(k+1)} + \hspace{-4mm} \sum_{n=-2,n\ne 0}^{N}\hspace{-3mm}  c^{(k+1)}_{mn}\mu^{(k+1)}_n K'_m(\mu^{(k+1)}_n R^{(k+1)}) B_{n\ell}^{(k+1)}.
\end{array}
\end{eqnarray}

  \item 
No flux conditions at the surface of the cylinder ($N+1$ equations):

\begin{eqnarray}
\begin{array}{l}
\displaystyle b_{m0}^{(K+1)}\mu^{(K+1)}_0J'_m(\mu^{(K+1)}_0 )B_{0\ell}^{(K+1)}+ \hspace{-4mm} \sum_{n=-2,n\ne 0}^{N}\hspace{-3mm}b^{(K+1)}_{mn}\mu^{(K+1)}_n I'_m(\mu^{(K+1)}_n) B_{n\ell}^{(K+1)} \\
\displaystyle +c_{m0}^{(K+1)}\mu^{(K+1)}_0H_m^{(1)'}(\mu^{(K+1)}_0 )B_{0\ell}^{(K+1)} + \hspace{-4mm} \sum_{n=-2,n\ne 0}^{N}\hspace{-3mm}c^{(K+1)}_{mn}\mu^{(K+1)}_n K'_m(\mu^{(K+1)}_n ) B_{n\ell}^{(K+1)}\\
\displaystyle \quad =0.
\end{array}
\end{eqnarray}

  \item 
Matching conditions of the wave elevation between adjacent layers ($K-1$ equations):

\begin{eqnarray}
\begin{array}{l}
\displaystyle b_{m0}^{(k)}E_0^{(k)}J_m(\mu^{(k)}_0 R^{(k+1)})+ \hspace{-4mm} \sum_{n=-2,n\ne 0}^{N}\hspace{-3mm} b^{(i)}_{mn} E_n^{(k)}I_m(\mu^{(k)}_n R^{(k+1)})\\
\displaystyle +c_{m0}^{(k)}E_0^{(k)}H_m^{(1)}(\mu^{(k)}_0 R^{(k+1)}) +\hspace{-4mm} \sum_{n=-2,n\ne 0}^{N}\hspace{-3mm} c^{(k)}_{mn}E_n^{(k)} K_m(\mu^{(k)}_n R^{(k+1)}) =\\
\displaystyle \quad b_{m0}^{(k+1)}E_0^{(k+1)}J_m(\mu^{(k+1)}_0 R^{(k+1)})+ \hspace{-4mm} \sum_{n=-2,n\ne 0}^{N}\hspace{-3mm} b^{(k+1)}_{mn} E_n^{(k+1)}I_m(\mu^{(k+1)}_n R^{(k+1)})  \\
\displaystyle \quad +c_{m0}^{(k+1)}E_0^{(k+1)}H_m^{(1)}(\mu^{(k+1)}_0 R^{(k+1)}) +\hspace{-4mm} \sum_{n=-2,n\ne 0}^{N}\hspace{-3mm} c^{(k+1)}_{mn}E_n^{(k+1)} K_m(\mu^{(k+1)}_n R^{(k+1)}),
\end{array}
\end{eqnarray}
where
\begin{eqnarray}
\displaystyle E_n^{(k)}=\frac{1}{\beta^{(k)}\mu_n^{(k)4}-\alpha \gamma^{(k)}+1}.\nonumber
\end{eqnarray}

  \item 
Matching conditions of the radial derivative of the wave elevation between adjacent layers ($K-1$ equations):

\begin{eqnarray}
\begin{array}{l}
\displaystyle  b_{m0}^{(k)}\mu^{(k)}_0E_0^{(k)}J'_m(\mu^{(k)}_0 R^{(k+1)})+ \hspace{-4mm} \sum_{n=-2,n\ne 0}^{N}\hspace{-3mm} b^{(k)}_{mn} \mu^{(k)}_nE_n^{(k)}I'_m(\mu^{(k)}_n R^{(k+1)}) \\
\displaystyle +c_{m0}^{(k)}\mu^{(k)}_0E_0^{(k)}H_m^{(1)'}(\mu^{(k)}_0 R^{(k+1)}) +\hspace{-4mm} \sum_{n=-2,n\ne 0}^{N}\hspace{-3mm} c^{(k)}_{mn}\mu^{(k)}_nE_n^{(k)} K'_m(\mu^{(k)}_n R^{(k+1)}) =\\
\displaystyle \quad  b_{m0}^{(k+1)}\mu^{(k+1)}_0E_0^{(k+1)}J'_m(\mu^{(k+1)}_0 R^{(k+1)})+ \hspace{-4mm} \sum_{n=-2,n\ne 0}^{N}\hspace{-3mm} b^{(k+1)}_{mn}\mu^{(k+1)}_n E_n^{(k+1)}I'_m(\mu^{(k+1)}_n R^{(k+1)}) \\
\displaystyle \quad +c_{m0}^{(k+1)}\mu^{(k+1)}_0E_0^{(k+1)}H_m^{(1)'}(\mu^{(k+1)}_0 R^{(k+1)}) +\hspace{-4mm} \sum_{n=-2,n\ne 0}^{N}\hspace{-3mm} c^{(k+1)}_{mn}\mu^{(k+1)}_nE_n^{(k+1)} K'_m(\mu^{(k+1)}_n R^{(k+1)}).
\end{array}
\end{eqnarray}

  \item 
Matching conditions of the bending moment between adjacent layers ($K-1$ equations):

\begin{eqnarray}
\mathcal{M}_m^{(k)} (R^{(k+1)}) = \mathcal{M}_m^{(k+1)} (R^{(k+1)}).
\end{eqnarray}

  \item 
Matching conditions of the shear force between adjacent layers ($K-1$ equations):

\begin{eqnarray}
\mathcal{V}_m^{(k)} (R^{(k+1)}) = \mathcal{V}_m^{(k+1)} (R^{(k+1)}).
\end{eqnarray}

  \item 
Free-free beam conditions; zero bending moment and shear force (4 equations):

\begin{eqnarray}\left\{
\begin{array}{l}
\displaystyle \mathcal{M}_m^{(1)} (b) = 0, \\
\displaystyle \mathcal{M}_m^{(K)} (1) = 0, \\
\displaystyle \mathcal{V}_m^{(1)} (b) = 0, \\ 
\displaystyle  \mathcal{V}_m^{(K)} (1)= 0.  
\end{array}\right.
\end{eqnarray}

\end{itemize}




\end{document}